\documentstyle[prb,twocolumn,aps,graphics,epsf,rotate]{revtex} 

\begin{document}

\twocolumn[\hsize\textwidth\columnwidth\hsize\csname
@twocolumnfalse\endcsname

\draft
\preprint{}

\title{Ferromagnetism in Diluted Magnetic Semiconductor Heterojunction Systems}

\author{Byounghak Lee$^{1,2}$, T. Jungwirth$^{2,3}$, and A.H. MacDonald$^2$}
\address{$^1$Department of Physics, Indiana University,
Swain Hall W.\ 117, Bloomington IN 47405}
\address{$^2$Department of Physics, University of Texas at Austin,
Austin TX 78712}
\address{$^3$Institute of Physics ASCR,
Cukrovarnick\'a 10, 162 53 Praha 6, Czech Republic}

\date{\today}

\maketitle

\begin{abstract}

Diluted magnetic semiconductors (DMSs), in which magnetic elements are substituted for 
a small fraction of host elements in a semiconductor lattice, can become ferromagnetic 
when doped.  In this article we discuss the physics of DMS ferromagnetism in systems
with semiconductor heterojunctions.  We focus on the mechanism that cause 
magnetic and magnetoresistive properties to depend on doping profiles, defect 
distributions, gate voltage, and other system parameters that can in 
principle be engineered to yield desired results.   

\leftskip 2cm
\rightskip 2cm
\end{abstract}
%\pacs{\leftskip 2cm PACS number: 73.40.Hm,73.20.Dx}

\vskip2pc] 

\narrowtext

%%%%%%%%%%%%%%%%%%%%%%%%%%%%%%%%%%%%%%%%%%%%%%%%%%%%%%%%%%%%%%%%%%%%%%%%%%%%%

\section{Introduction}

In many semiconductor crystals, substitution of a transition metal 
element for a host element adds a local moment to the system's low-energy 
degrees of freedom.  Systems in which a transition element is substituted
quasi-randomly on a finite fraction, $x$, of a host semiconductor element sites
are known\cite{dmsreview} as diluted magnetic semiconductors (DMSs).
The degeneracy of each isolated moment's magnetic state manifold makes these 
materials very sensitive to the host material 
environment in which they are placed and to external perturbations.  
From a very general point of view, this circumstance is attractive for 
the development of electronic systems that have technologically useful 
properties.  
A ferromagnetic state, in which long range order is established among 
the local moments, is especially important for technology since the moments
then act collectively leading to greater sensitivity and more robust phenomena.  
This idea has been pursued for nearly twenty years.  
Over time it has been established that the behavior of substitutional 
Mn elements is particularly simple in many host semiconductors, because of 
a strong tendency toward well defined $S=5/2$, Mn$^{2+}$ local moments.  
In II-VI semiconductors,\cite{dmsreview} 
Mn substitutes on cation sites where it provides 
local moments but alters the host semiconductor valence and conduction
bands very weakly.
The Mn local moment and the conduction and valence band electrons are coupled
by on-site exchange interactions, due to direct overlap between sp and d orbitals 
and to hybridization between d and p orbitals.\cite{larsonprb88}  
The local moments are easily aligned by relatively weak magnetic fields, 
leading to a rich variety of magnetooptical effects.  
These materials are not ferromagnetic however.  
In III-V compound semiconductors on the other hand, substituted 
Mn$^{2+}$ ions both lead to local moment formation and act as acceptors, 
introducing valence band holes and adding to the degrees of freedom that 
are important in determining the system's electronic properties. 
The holes mediate interactions between the Mn moments, correlating their 
orientations and making ferromagnetism possible.  

Since Mn is not soluble in III-V semiconductors, it can be incorporated 
only by non-equilibrium growth techniques.
The discovery of ferromagnetism\cite{classicrefs} in low-temperature MBE grown 
(In,Mn)As, and the eventual realization\cite{ohnogamnas,ohnosci98} that relatively high 
ferromagnetic transition temperatures 
could be achieved in (Ga,Mn)As, if the samples were 
suitably annealed after growth, has led to an explosion of interest.  
Room temperature ferromagnetism has been 
reported\cite{sonoda,reedapl01,theodoropoulou02,chen} in several 
different Mn-doped DMS materials very recently.  Equally exciting is a recent
report\cite{parksci02} that ferromagnetism occurs in DMS's based on the the 
group IV element, Ge.  The story is still unfolding as these words are being written;
annoucements of new surprises have been appearing on e-print servers and in research journals 
with startling regularity.

An aspect of the ferromagnetism that is key for many potential applications is 
that the {\em itinerant holes in the valence bands are full 
participants.} The electronic structure of the quasiparticles that carry 
charge through the system and control its optical absorption and luminescence, 
are very sensitive to the magnetic state of the system. 
Considerable progress has been made in understanding the magnetic, transport,
and optical properties\cite{dietlprb01,ourreview} of bulk DMS ferromagnets, 
although current understanding is certainly far from complete.  
A  thorough review of this research can be found elsewhere\cite{dietlhere} 
in this issue of Semiconductor Science and Technology.
An interactive web resource that provides information,
for many different bulk and heterojunction DMS ferromagnet systems, 
on predictions made by the semi-phenomenological models we discuss below
is available at http://unix12.fzu.cz/ms.

The purpose of this article is to address the physics of ferromagnetism in 
DMS ferromagnet heterojunction systems, 
a research topic that is far from mature and, in fact, relatively unexplored 
at present.  It would appear that the band-structure engineering\cite{nobelprize} 
possibilities of heterojunction systems should open up unparalleled opportunities 
for tuning magnetic, magnetooptic, and magnetotransport properties of these 
low-dimensional and layered ferromagnets.  
These properties are likely to have technological impact only to the extent 
that ferromagnetism, or at least strongly enhanced paramagnetism, can be achieved at 
room temperature.  It will likely be some years before materials science 
research motivated by the search for high-transition-temperature low-carrier-density 
DMS ferromagnets becomes mature.  We attempt here to provide
some guidance, based on simple theoretical considerations, that we hope will
be helpful for what we expect to be an ongoing research effort.
The scope of this article is limited by our incomplete 
perspective on this rapidly developing field; the topics on which
emphasis has been placed are those on which we have worked ourselves.

The article is organized as follows.    
In Section II we introduce the semi-phenomenological model of DMS
ferromagnets on which our discussion is based, and address the 
physics that controls the ferromagnetic critical temperature.   
In Section III, we review
current experiments and theoretical predictions for different 
DMS heterojunction systems, pointing out research directions that 
might be helpful in sorting out the interplay of materials and fundamental
physics issues that will need to be resolved to make progress. 
Section IV contains a summary of this article.  
Although we discuss only magnetic and dc transport properties here, 
the magnetooptical properties of DMS heterojunction ferromagnets will likely 
prove to be an equally exciting field of research over the coming decade.  
We restrict our attention in this article to the physics of ferromagnetism 
in DMS systems composed of different semiconductors stacked in layers formed 
by following an epitaxial growth protocol.  

%%%%%%%%%%%%%%%%%%%%%%%%%%%%%%%%%%%%%%%%%%%%%%%%%%%%%%%%%%%%%%%%%%%%%%%%%%

\section{Remarks on models and theories of DMS ferromagnetism} 

\subsection{Semi-phenomenological model}

We discuss DMS ferromagnets here mainly in terms of a widely accepted 
semi-phenomenological model that has three terms:
\begin{itemize} 

\item The coupling of the $S=5/2$ Mn spins to the external magnetic field,
$  g \mu_B \sum_I \vec S_{I} \cdot \vec H_{ext}$
.

\item The Hamiltonian of the host  semiconductor valence bands, 
described using a multi-band envelope function formalism.\cite{laserbook}
{\em For many properties it is absolutely essential 
to incorporate spin-orbit coupling in a realistic way.}  
Six or eight band model that includes the `split-off' band and/or 
the conduction band are sometimes desirable.   
This band Hamiltonian should include the effect of strain due to lattice 
matching between the epitaxially grown (II,Mn)VI, (III,Mn)V or (IV,Mn) films
and the substrates on which they are grown.    

\item Antiferromagnetic exchange coupling between the $S=5/2$ Mn$^{2+}$ spins 
and the valence band holes, $ J_{pd} \sum_I \vec S_{I} \cdot \vec s(\vec R_{I})$.  
This interaction represents virtual coupling to states that have been 
{\em integrated out} of the model's Hilbert space, in which electrons 
are exchanged between the Mn ion d shells and the valence 
band.\cite{dmsreview,larsonprb88}
These exchange interactions are isotropic to a good approximation because the 
Mn$^{2+}$ ion has total angular momentum $L=0$. 
\end{itemize}

The most fundamental assumption of this model is that the Mn $d$ electrons 
are localized by their strong Coulomb interactions.  
The low-energy 
degrees of freedom are then valence band holes and one $S=5/2$ local moment for 
each Mn ion; other terms that involve these degrees of freedom can be 
added to the Hamiltonian if they are believed to be important.
For example, the present discussion neglects Coulomb interactions between holes and 
Mn acceptors, and short-range antiferromagnetic interactions between Mn spins, 
among other terms,\cite{ourreview} that can be of 
importance\cite{yang:mit} in the most general circumstances. Since we are 
interested here in the strongly metallic limit,  Coulomb interactions are not 
expected to play an essential role in the ferromagnetism.  
We will also  make the approximation that the Mn ions can be replaced 
by a continuum density that can have a prescribed dependence on the growth 
direction coordinate, $N_{Mn}(z)$.  
This approximation has the advantage of removing disorder in the direction 
perpendicular to growth (due to randomness in Mn ion locations) from the model, 
profoundly simplifying its analysis.  
Disorder plays an essential role, of course, in determining the magnitude 
of transport coefficients.  
For many purposes, however, the continuum Mn density 
approximation can be a reasonable one provided 
that the typical distance between Mn ions is smaller than other characteristic 
lengths such as the average distance between valence band holes, the Bohr radius,
or quantum well widths. 
The utility of this approximation is often improved by the fact that 
compensation of Mn acceptors by antisite defects is present in all these 
DMS ferromagnets.  
From a microscopic point of view, neglect of Mn ion site disorder is 
equivalent to a virtual crystal approximation.  

In using a model with one $S=5/2$ local moment per Mn and valence band holes
to discuss DMS ferromagnetism we are following {\em Occam's razor}.  Although
this model is well established by experiment in the most heavily studied bulk
systems, (Ga,Mn)As and (Ga,In)As, it may not apply to all bulk systems; possibly
not to nitrides for example\cite{dietlhere}.  It also may not fully
capture the physics of systems in which it is common for Mn ions to be 
in neighboring positions, for example in the $\delta$-doped layers discussed
below.  It also could turn out that the most useful DMS ferromagnets 
will ultimately be created by intentional clustering of magnetic elements that 
invalidates our model's assumptions.
At this stage however, it seems appropriate to determine how much of 
the rich behavior of these electrons can be understood assuming that the Mn d-electrons are 
always localized.  The alert reader should keep in mind
the possibility that the model on which the theoretical discussion in this 
article is based may not always apply.

A useful zeroth order picture\cite{ourmftpaper,dietlprb01,ourreview} 
of DMS ferromagnets follows 
from combining the Mn-continuum (virtual crystal) approximation with 
a mean-field treatment of the kinetic exchange interaction.  
The mean-field treatment does not allow any fluctuations in 
the magnitude or direction of Mn ion or valence band hole spin polarizations.  
In this picture each Mn ion is described by a mean-field Hamiltonian 
$   \sum_I \vec S_{I} \cdot  [  g \mu_B \vec H_{ext} + J_0(z) ]$ 
where $J_0(z) =  J_{pd} {\vec s}(z)$ and ${\vec s}(z)$ is the valence band 
spin polarization.
$J_0$ is an effective field seen by the local moments due to spin-polarization 
of the band holes, analogous to the nuclear Knight shift.  
Similarly the band holes have an additional mean-field term in their 
Hamiltonian: $\sum_i \vec s_i \cdot \vec h (z)$, where $\vec s_i$ is 
the i-th hole spin, $\vec h (z) = J_{pd} N_{Mn}(z) \langle\vec S\rangle
 (z)$ is 
an effective magnetic field experienced by the valence band holes and 
$\langle\vec S\rangle (z)$ is the mean 
spin polarization of the Mn spins.  This quantity is 
given\cite{ourreview} in magnitude and 
direction  
by the free-spin Brillouin function expression using 
the total mean field, $\vec H_{eff}=\vec H_{ext}+J_0/g\mu_B$.  
The hole spin polarization,
${\vec s}(z)$, is determined by solving the hole Schroedinger equation 
with a space-dependent external Zeeman field.   This self-consistency 
condition closes the mean-field equations,\cite{ourmftpaper} 
and allows any artificial spatial
structure to be described.
 
The value of $J_{pd}$ in (Ga,Mn)As is 
approximately\cite{okabayashiprb98,omiyaphyse00} 
$60 {\rm meV}{\rm nm}^3$.
This parameter is not expected to vary widely from material to material, 
so we can use this value to get a feeling for the  strength of these 
mean-fields in different circumstances. 
For 5 \% Mn substitution $x=0.05$, the density of Mn ions is $N_{Mn} 
\approx 1 {\rm nm}^{-3}$.  
At present the highest values of $T_c$ tend to occur near this value of $x$.   
It is not
yet known whether or not there is a material strategy that will avoid the
tendency of $T_c$ to decrease at larger value of $x$. 
The density of valence band holes in the bulk for this case has 
been measured\cite{omiyaphyse00} to be 
$p\approx 0.35 {\rm nm}^{-3}$, 
although this quantity is expected to be 
sensitive to growth temperature and annealing procedure.
Using these values we obtain a characteristic energy scales of the Mn spin 
mean-field at complete hole spin-polarization which is $ J_0\sim 10 {\rm meV}$.
Similarly, the valence band mean Zeeman-field at full Mn spin polarization is 
$h\sim 150 {\rm meV}$. 
{\em We note that the band spin-splitting field is comparable to 
the Fermi energy, $\epsilon_F$.  
This important property explains why transport, optical and other 
quasiparticle properties of DMS ferromagnets are extremely sensitive
to the system's magnetic state.}  

\subsection{Comments on simple $T_c$ estimates}

The most important property of a DMS ferromagnet, from a practical point of view,
is its ferromagnetic transition
temperature $T_c$.  With this in mind, it is useful to discuss prospects and strategies for 
obtaining large values of this parameter in DMS heterojunction systems.  
An important objective of this paper is to provide some theoretical guidance 
for efforts to achieve high $T_c$ values in delta-doped and quantum well 
systems.  
It turns out\cite{dietlprb01,ourreview,jungwirth02} that in bulk systems, 
mean-field theory implies a critical temperature 
$ T_c \sim  h J_0/ k_B \epsilon_F \sim S J_0/k_B \sim 100 {\rm K}$.
We will argue below that both the mean-field expression and $S J_0/k_B$ provide 
loose upper bounds for $T_c$ and that both parameters must be made large if 
high ferromagnetic transition temperatures are to be achieved.

Several different arguments can be given to estimate the ferromagnetic
critical temperature of DMS ferromagnets.  
While each is useful, it is important to understand their limitations, 
especially when applying them to the wide variety of different circumstances 
that can be realized in heterojunction systems.  
Mean-field theory has been a very useful guide for bulk systems.  
The mean-field theory\cite{ourmftpaper,dietlhere} critical temperature 
$T_c^{MF}$ is the temperature below which the free energy 
can be lowered by creating uniform oppositely directed spin-polarizations 
in Mn ion and valence-band-hole subsystems.
Spin-polarization results in a reduction of entropy of the Mn system, and 
an increase in the band energy of the valence band, and a compensation gain 
in kinetic exchange energy.  
For a single-parabolic band 
$k_B T_c^{MF} =  (SJ_0 h/\epsilon_F) (1+1/S)/4$.  
Using $J_0 \sim 10 {\rm meV}$, $h \sim 150 {\rm meV}$ and 
$\epsilon_F \sim 100 {\rm meV}$, gives $T_c \sim 100K$ for typical 
bulk $T_c$ estimates.  
This formula gives a fairly solid upper bound on the critical temperature, 
and certainly allows for room temperature 
ferromagnetism under favorable circumstances,
but can overestimate $T_c$ by a larger fraction in some cases. 

A second loose upper limit on the critical temperature is especially relevant to 
quantum well and other heterojunction systems.  It follows from noting 
that $J_0$ is an upper bound on the cost in energy for a single Mn spin to flip in 
a fully ferromagnetic background.  The fact that $J_0$ overestimates the cost of an
uncorrelated spin-flip follows from a variational argument.  If the electronic 
system state is held fixed, the change in energy when a Mn spin orientation 
is changed, reducing its projection 
along the order direction  by one unit, is equal to the expectation 
value of the change in the Hamiltonian operator, {\it i.e.} to $J_0 >0$.  The actual
change in ground state energy will be smaller, since the electronic system 
will respond to the change in the Hamiltonian and the new ground state will
have a lower\cite{ourreview,konigprl00}, energy.
The entropy gain from uncorrelated spin fluctuations will destroy ferromagnetism
when this energy cost is too small compared to the thermal energy $k_B T$.  
A loose $T_c$ upper bound is $k_B T_c^{SF} = S J_0 = S J_{pd} p /2$, 
which will be smaller than $T_c^{MF}$ in many heterojunction systems.  
Using this $T_c$ bound, {\em it follows that ferromagnetic transition temperatures 
in excess of $100 {\rm K}$ are impossible, given our model and the 
estimate we use for $J_{pd}$, unless the 3D carrier density exceeds 
$\sim 0.1 {\rm nm}^{-3}$.}  
It follows that high ferromagnetic transition temperatures will not be 
possible in modulation doped II-VI quantum well systems where 2D carrier 
densities are limited by electrostatic and band-alignment considerations to 
$~\sim 10^{12} {\rm cm}^{-2} = 10^{-2} {\rm nm}^{-2}$, even when high transition
temperatures are predicted by mean-field theory.

The typical energy required to flip a single Mn spin 
can be substantially smaller than $J_0$, if the hole system response 
is strong.  This strong response occurs when the semiconductor band 
dispersion is weak\cite{konigprl00,schliemannapl01}
or when the band electrons are strongly localized\cite{millis01,schliemannprl02}; in
both cases the energetic cost for a response in band-electron spin orientation
that maintains the full kinetic exchange energy is small, so that the 
overall cost of spin reorientations is reduced substantially below $J_0$, 
reducing the ferromagnetic critical temperature even below this estimate.

Disorder can also have an important effect on the critical 
temperature.\cite{schliemannprl02,disorder,dassarma02,ourreview}
When disorder is included, the hole density tends to increase at Mn spin sites,
and $J_0$ increases proportionately.
The same peak in hole density tends, however, to decrease coupling between 
Mn spins.  It appears likely that the net effect of disorder is always to decrease 
the maximum temperature at which long range magnetic order can be 
maintained, although it has not been established that this is the case.
{\em Higher ferromagnetic transition temperatures will tend to occur in 
heterojunction systems in which Mn spins are coupled by more weakly disordered 
valence band holes.} 

\section{DMS Ferromagnet Heterojunction Systems} 

In this section we distinguish and discuss separately four
classes of systems; i) layered DMS ferromagnet thin films in which the 
semiconductor layers are thick enough that quantum confinement effects
within each layer do not play an important role in the physics,
ii) quantum wells in which quantum confinement effects plays a central role 
and only one or two subbands are occupied, 
iii) delta-doped layers in which a high concentration of Mn acceptors is placed
at specific growth direction coordinates, resulting in the occupation of several
growth direction subbands, 
and iv) superlattices that consist
of alternating magnetic and non-magnetic layers thin enough for quantum confinement
to play an essential role.  The artificial spatial structure introduced by semiconductor 
heterojunctions plays different roles in these different cases.
Although we find this classification scheme useful, there are obviously grey areas 
between these four cases which can be usefully explored.

\subsection{Layered DMS Ferromagnet Thin Films}

We first discuss the case of thick films in which
quantum-size effects play no important role in determining magnetic 
properties like the ferromagnetic transition temperature. 
Because carrier densities in these semiconductor 
ferromagnets are lower than in metals, quantum size effects will start to become important 
at larger film thicknesses.
We will argue below that it will not be possible to obtain
large $T_c$'s in DMS ferromagnets unless the typical 3D carrier density
$p$ is comparable to that in the high $T_c$ bulk samples.  
The number $I$ of occupied 2D subbands in a film of width $w$ can be roughly 
estimated for semiconductors using the particle-in-a-box expression 
$I^2 \sim p w^3$.  
For $p \sim 0.3 {\rm nm}^{-3}$, $I$ is larger than 15, and we can safely
assume properties similar to those of the bulk, for $w$ larger 
$\sim 10 {\rm nm}$.
In this section we have in mind heterojunction systems with ferromagnetic layer
thicknesses in excess of this value.    

Magnetoresistance effects in layered structures of magnetic and non-magnetic 
metals have been studied extensively, following the discovery of the giant 
magnetoresistance effect in magnetic multilayers.  
The basic physics is simplest and the opportunities for applications are 
greatest in the  {\em spin valve} structure, in which two ferromagnetic 
metals are separated by a non-magnetic metal.  
The role of the non-magnetic metal is to weaken the exchange coupling between 
the two ferromagnets to a negligible value so that the magnetizations of the 
two magnetic layers can be manipulated independently.  
The resistance for current flow between the two magnetic layers tends to be 
substantially larger when the their magnetizations have opposite orientations.
The fundamental origin of the effect is the spin-dependent effective potential 
experienced by the quasiparticles that carry current 
through the system because of their 
{\em exchange} interactions with other quasiparticles 
in the ferromagnetic state.
When the magnetization configuration changes, the spin-dependent effective 
potential changes, and all quasiparticle properties of the metal change.  
A similar and actually somewhat simpler effect known as tunnel 
magnetoresistance occurs when the ferromagnetic metals are separated by an insulating 
barrier, allowing transport to be described using a tunneling Hamiltonian 
formalism.  
As emphasized above, the current-carrying quasiparticle states of DMS 
ferromagnets also experience very strong spin-dependent effective potentials,
due to exchange interactions with both local moments and other quasiparticles, 
that are sensitive to their magnetic configuration. 
{\em We should therefore expect that very strong magnetoresistance effects will 
occur in layered DMS ferromagnet thin film systems}.  This property  already 
has clearly been established in experiment.

DMS Ferromagnet/ Nonmagnetic semiconductor/ DMS Ferromagnet trilayers have been 
prepared and studied by the Ohno {\em et al.}\cite{ohnojmmm99} 
and Tanaka {\em et al.}\cite{tanakatmr}.
The (Ga,Mn)As magnetic layer thicknesses were $\sim 30 {\rm nm}$ in the Ohno 
group case and $\sim 50 {\rm nm}$ in the Tanaka group case, both safely in 
the bulk film limit.
Ohno {\em et al.} samples had (Al,Ga)As barrier layers with a relatively small 
Al fraction, while the Tanaka {\em et al.}  grew pure AlAs barriers with 
a range of thicknesses.  
By growing magnetic layers with different Mn fractions, these authors
were able to make the coercive fields of the two magnetic layers distinct.  
Growth direction resistance in these samples shows a large increase over 
the relatively narrow field range where the magnetization has reversed in 
one film but not the other, as expected from the theories of giant 
magnetoresistance and tunnel magnetoresistance and the strong spin-polarization 
of DMS ferromagnet valence bands at low-temperatures.
In the Tanaka {\em et al.} samples, 
relative changes in the growth direction resistance
(the tunnel magnetoresistance ratio) approaching 80 \% were obtained at 
low temperatures for $\sim 1.5 {\rm nm}$ thick AlAs barrier layers.  
Using a simple parabolic band model, the authors attribute the dependence 
of the TMR ratio on barrier thickness to the combined effect of differences 
between majority and minority spin Fermi radii and the dependence of tunneling 
amplitude on growth direction velocity.  
These results demonstrate that it will be possible to achieve very high 
resistance ratios in DMS ferromagnet trilayer structures.  
In addition it should prove possible to tune between tunnel magnetoresistance 
(TMR) and spin-valve regimes by adjusting the height and width of the barrier 
both chemically, by varying the Al content in the barrier for example, or 
by doping the barrier with non-magnetic acceptors like Be to reduce its 
electrostatic height.

A layered structure that has been studied extensively in non-magnetic 
bulk semiconductors is the double-barrier resonant tunneling diode.  
Ohno and collaborators have studied\cite{ohnosci98,ohnojmmm99} 
the quasiparticle states of bulk
DMS ferromagnets by examining the current-voltage characteristics of 
resonant-tunneling diodes with ferromagnetic emitters.   
{\em These experiments are important because they establish experimentally 
two key features of the ferromagnetic state.} 
First they established the approximate reliability of 
the mean-field picture of ferromagnetism in DMS ferromagnets because 
they are able to follow the temperature dependence of the splitting of features 
in the semiconductor bands that are, at least approximately, proportional to 
the band effective Zeeman field $h$.  
The experiments show that $h$ approaches zero near the $T_c$ value 
established by bulk magnetization measurements.  
If mean-field theory were not applicable, the bands would be expected to be 
locally and instantaneously spin polarized at temperatures much larger than 
the thermodynamic critical temperature, which is the temperature at which long range 
magnetic order is established.  
The demonstration of band spin-splitting is also an experimental proof that 
the quasiparticle bands are full participants in the magnetic state, as 
implied by the mean-field theory picture.  
The quasiparticle bands do indeed experience a spin-dependent 
effective potential in the magnetic state. 
The demonstration that these splittings vanish at the critical temperature 
provides important guidance for the development of reliable theoretical
descriptions of these materials, since the fact that mean-field theory
is relatively accurate, at least for $x \sim 0.05$ high-$T_c$ (Ga,Mn)As,
is still not universally accepted.

A unique property of DMS thin films was demonstrated by 
Ohno {\em et al.}\cite{ohnonature00} who used a field-effect transistor
structure to vary reversibly the ferromagnetic transition temperature by 
applying external electric field. The magnetic (In,Mn)As layer in this system is very thin,
only 5~nm, so that the sample actually lies at the border between 3D thin films and 2D 
quantum wells. When approached from the 3D side, a part of the $T_c$ dependence
on bias can be attributed to a change in the carrier density induced
by the applied electric field. Another contribution to the effect can
be understood from the quantum well point of 
view\cite{hauryprl97,leeprb00,breyprl00,lee01} 
in which the overlap
between the growth-direction wavefunction and the Mn-doped region
increases or decreases with applied bias, enhancing or suppressing
ferromagnetism in the magnetic layer. The latter contribution belongs to the class
of quantum confinement effects that are essential for quantum wells with one or
few occupied subbands, a condition easily achieved in modulation doped (II,Mn)VI
quantum wells that we discuss next.

Finally we mention the possibility of completely altering the magnetic anisotropy of 
DMS ferromagnet thin films by changing the substrate on which they are grown,
a property which was already established by early experiments\cite{classicrefs,ohnogamnas}.
The origin of magnetic anisotropy in DMS ferromagnets is the strong spin-orbit
interactions in semiconductor valence bands.  Spin-orbit interactions have a 
strong influence on details of the coupling between Mn moments\cite{janko}
and on the energy of correlated Mn spin-fluctuations\cite{konigspinwave}, 
but are fully responsible for magnetic anisotropy.  The systematics of the 
dependence of magnetic easy axis on substrate, appear to be fully 
explained\cite{dietlprb01,abolfathprb01} by including the influence of 
lattice-matching strains on the host valence band electronic structure.
These strains lift the cubic symmetry of the semiconductor and in so doing
dominate the magnetic anisotropy.

\subsection{Quantum Wells} 

Ferromagnetism in modulation doped II-VI DMS semiconductor quantum well
systems was predicted by Dietl et al.\cite{hauryprl97} and 
confirmed by Kossacki {\em et al.}.\cite{kossackiphyse00}.  
For the case of a parabolic band, the mean-field-theory expression for 
the critical temperature is a simple generalization of its bulk 
counterpart:\cite{hauryprl97,leeprb00}
\begin{equation} 
T_c^{MF} = {S (S+1) \over 12} {J_{pd}^2 \over k_B} {m^{\ast} \over \pi \hbar^2}
\int dz |\psi(z)|^4 N_{Mn}(z) \:.
\end{equation}
where $N_{Mn}(z)$ is the Mn density distribution in the quantum well, 
$m^{\ast}$ is the 2D band mass, and $\psi(z)$ is the occupied subband 
wavefunction.   
This expression suggests possibilities for manipulating magnetic properties 
with gates that are an important part of what is interesting about these ferromagnets.

\begin{figure}
\epsfxsize=3.5in
\centerline{\epsffile{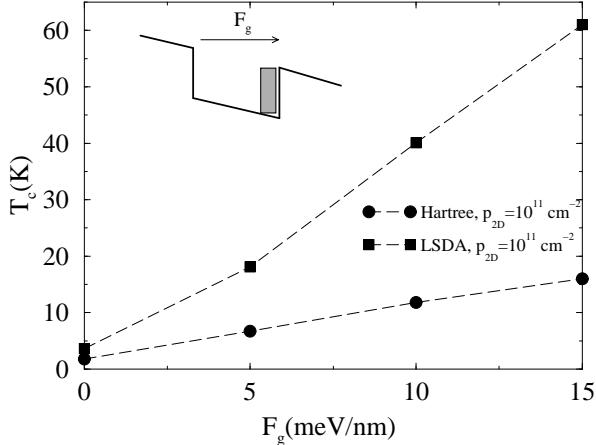}}
\caption{
Dependence of $T_c$ on bias voltage applied across  the
$w=10$~nm wide quantum well partially occupied
by magnetic ions with $N_{Mn}=1$~nm$^{-3}$
over a distance of 3~nm near one edge of the quantum well.
Circles (squares) represent results of the full numerical self-consistent
calculations without (with) the exchange-correlation potential.
}
\label{tcfg} 
\end{figure}

In Fig.~\ref{tcfg} we plot mean-field theory $T_c$ results we 
calculated\cite{leeprb00} for a quantum well that is 
partially filled with Mn ions with a density of $1 {\rm nm}^{-3}$, 
using a heavy-hole band in-plane mass $m^{\ast} = 0.11 m_0$.
The figure illustrates the increase of $T_c$ with an applied
bias voltage which draws electrons into the magnetic ion region, 
an effect mentioned in the previous section.
As also emphasized in this figure, when hole-hole interactions are included
in mean-field theory, $T_c$ estimates are enhanced, especially at lower
2D densities.  
The situation illustrated here is fairly typical of potential 
modulation doped II-VI semiconductor DMS ferromagnets.  

It is important to mention at this point that mean-field 
$T_c$ estimates can be too high in quantum well systems.
For the parameters used to obtain Fig.~\ref{tcfg}, at a typical point in the 
quantum well $J_0 = J_{pd} p_{2D} / w \sim 0.016 {\rm meV}$,
whereas $T_c^{MF} \sim 1 {\rm meV}$ 
(neglecting interactions).
We have argued above that $T_c$ cannot exceed $S J_0 \sim 0.04 {\rm meV}$, 
so in retrospect it seems unlikely that mean-field theory estimates are reliable
for the case illustrated.  Using $J_{pd}$ values closer to those now 
believed to be accurate reduces the discrepancy between mean-field-theory $T_c$ 
estimates and this approximate bound, reducing the former quantity by a 
factor $\sim 10$ and the later by a factor $\sim 3$, but also reduces 
the predicted critical temperatures to less interesting values. 
These caveats do not alter the basic point that magnetic and other properties of 
DMS quantum well ferromagnets will be particularly easy to control, but they
do have implications for strategies to achieve high $T_c$ values, as we
discuss below.
 
In the quantum well systems studied experimentally\cite{kossackiphyse00} by
Kossacki {\it et al.}, a critical temperature 
$\approx 2 {\rm K}$ was extracted by looking for spontaneous spin-splitting of 
PL features.  In this case the measured $T_c$ is in a good 
agreement with mean-field theory
estimates predictions based on the smaller values of $J_{pd}$ favored 
by more recent experiments.  
The quantum well widths in these samples are narrower 
than for the situation illustrated above  so that the mean-field theory 
is also reasonably compatible with the $SJ_0$ bound in this case.  

Both mean-field theory and the $SJ_0$ bound should be considered in 
devising strategies for raising the ferromagnetic critical 
temperature.  
For example, mean-field-theory predicts very large transition temperatures 
when the normal-state density-of-states is extremely large, as often happens 
in valence-band quantum-well systems at densities slightly larger than 
those at which second subband is first occupied.
In our view, the $SJ_0$ bound will place a severe limit on the $T_c$ 
increase that can be achieved in this way. 
$T_c$ values much higher than those obtained to date will require 
narrower quantum wells and higher 2D hole gas densities. 

In general the mean-field theory is expected to provide more
reliable predictions for the ground-state
and low temperature properties of DMS heterojunctions on which we 
focus of the remaining parts of this article.  
As we now explain, low density II-VI DMS quantum wells can have very
strong and tunable magnetic anisotropy. The interplay between local moment 
- itinerant hole exchange coupling,
Zeeman coupling of the localized moments to external magnetic field, and
the coupling of an external electric field to the orbital degrees of freedom
of itinerant holes, makes it possible to vary the coercive field
by over an order of magnitude with rather modest external electric fields.
As a result, the mean-field theory predicts\cite{lee01} 
that the magnetization orientation in quantum
wells can be manipulated electrically without changing the
magnetic field.
Our calculations also suggest that capacitance measurements can be used to
probe the magnetic state of biased ferromagnetic semiconductor quantum wells.

\begin{figure}[tb]
\epsfxsize=3.5in
\centerline{\epsffile{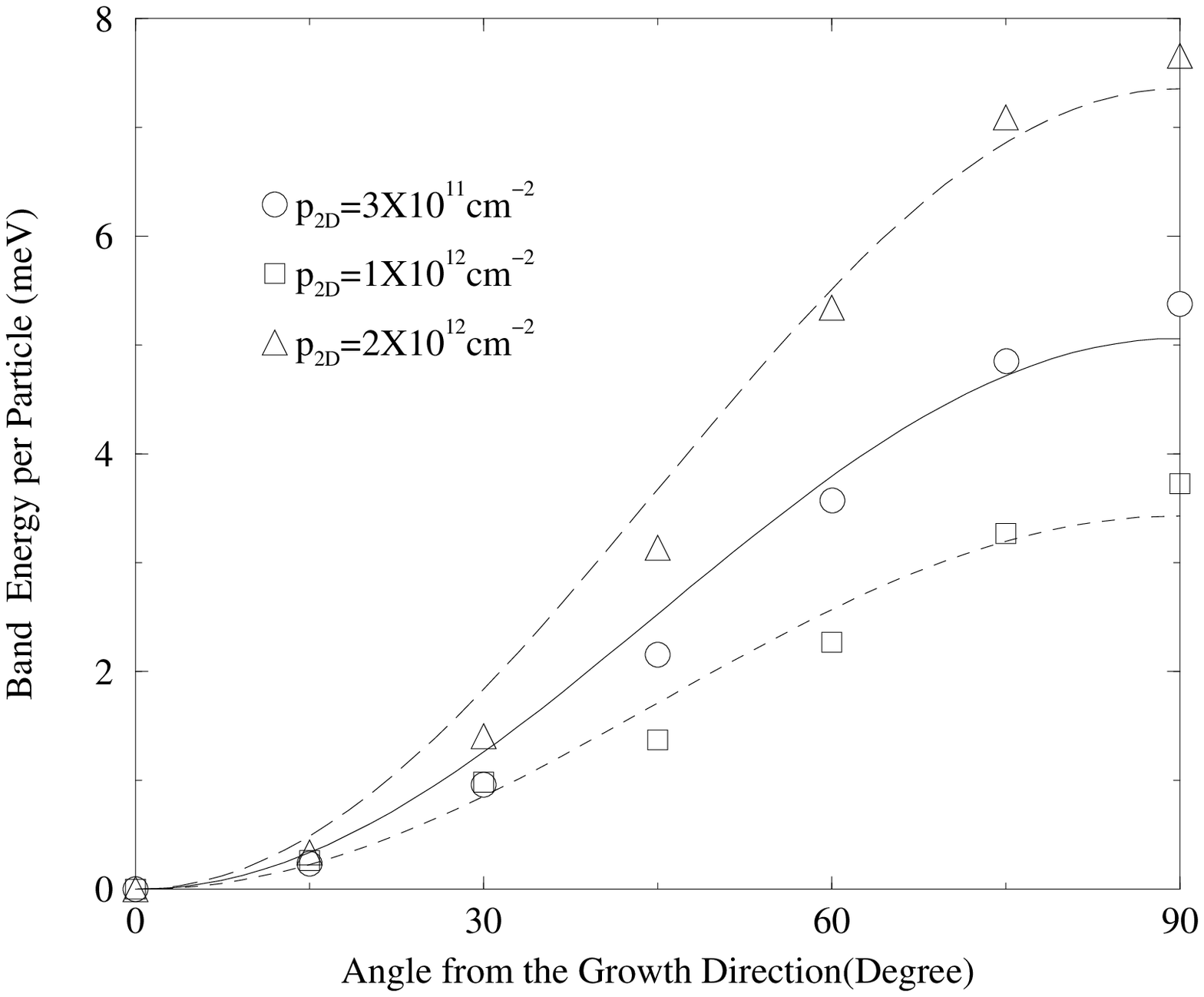}}
\vspace*{0.0cm}
\caption{Magnetic anisotropy energy calculated from the
self-consistent Hartree
approximation using the four-band Luttinger model
as a function of magnetization direction from the growth direction.
The magnetic anisotropy results from the combined effects of 
spin-orbit interactions in the valence band and quantum confinement.
These plots are for a (Cd,Mn)Te/(Cd,Zn)Te quantum well of the width
$w = 10$~nm, with a valence
band offset of 150~meV and uniformly distributed Mn ions of density
$N_{Mn}=6\times10^{20}$~cm$^{-3}$.
The anisotropy energy is defined relative to the energy 
for magnetization oriented along the growth direction orientation,
which is the easy axis according to these calculations. 
At the lowest density only one subband is occupied while two subbands are
occupied at the two higher densities.
The curves fit the numerical data using the usual phenomenological expression for 
uniaxial magnetic anisotropy.}
\label{anisoqw}
\end{figure}

The magnetic anisotropy energies plotted in Fig.\ref{anisoqw} were obtained
using the four-band Kohn-Luttinger model that describes only the total
angular momentum $j=3/2$ bands, and is adequate at low hole densities
when spin-orbit coupling is large. 
The degeneracy between heavy-hole ($|j_z|=3/2$) and light-hole
($|j_z|=1/2$) bands at the $\Gamma$-point in the bulk is lifted by size
quantization effects in a quasi-two-dimensional system.
The resulting heavy-light gap is larger than the Fermi energy, 
in the relevant range of hole
densities, allowing only the two heavy-hole bands to be occupied.
The heavy holes have their spin aligned along $\hat{z}$-axis
($\langle 001\rangle$ crystal direction) so that the band electron spin
matrix elements get smaller when the magnetization tilts away from
the growth direction.   This leads to
very strong magnetic anisotropy with easy axes along and opposite to
the growth direction.  This anisotropy is reduced in magnitude but is
still large, as seen in Fig.\ref{anisoqw},  when mixing between
heavy- and light-hole bands is fully accounted for. Note that in DMS
quantum wells, the magnetic anisotropy per itinerant carrier 
is several times larger than
in bulk DMS\cite{abolfathprb01,dietlprb01} systems and nearly three orders of
magnitude larger than in cubic transition metal ferromagnets.\cite{skomski}

\begin{figure}[tb]
\epsfxsize=3.5in
\centerline{\epsffile{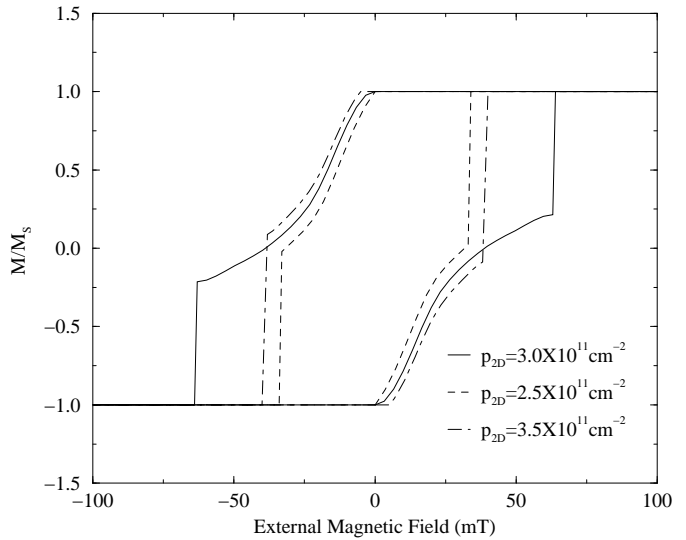}}
\vspace*{0.0cm}
\caption{Local moment hysteresis loops in a quasi-2D DMS system with
different carrier densities.
These curves were obtained by using mean-field theory with a Hartree approximation
for hole-hole interactions.  The coercive field is sensitive to a gate voltage
that alters the hole density in the quantum well because of interplay between
quantum confinement and quasiparticle-state spin-splitting.} 
\label{hys}
\end{figure}

We have predicted that 
the magnetization reversal process is quite unusual in DMS quantum
wells.
When the magnetic field is applied, direct Zeeman coupling to a local
moment competes with the local mean-field
kinetic-exchange coupling, which is proportional to the local 
itinerant-hole spin-density.
Since the carrier density is smaller at the edges of the quantum well
than at the center, spin reversal starts from the well edges.
This,  in turn,  creates an exchange barrier for the majority-spins
which effectively narrows the quantum well in which they reside.
At the same time an effective
double-well potential, attractive at the outer edges of the 
quantum well, develops for the minority spins.
As the external magnetic field increases,
the minority-spin energy levels are lowered and
the majority-spin energy levels are raised.
When the lowest down-spin energy level reaches the Fermi energy,
holes start to occupy the down-spin states.  Our
self-consistent calculations show that once this happens,
the magnetization reversal is rapidly completed and only the
uniform down-spin state is stable.
Fig.\ref{hys} illustrates the possibility of modifying hysteresis loops
electrically which, in turn, allows the magnetization to be changed by the
external electric field. This attractive feature of ferromagnetic semiconductor
quantum well systems, somewhat reminiscent of recent
interest in current driven magnetization
reversal\cite{curswitch} in metallic ferromagnets, is only a theoretical
prediction at present.  We expect that the true picture is more complex, but
believe that this calculation illustrates one rather unique aspect of DMS 
ferromagnets that should be accessible to experimental study, once gateable 
systems can be reliably grown.  This behavior originates from the fact that 
spin-splitting energies experienced by holes in a quantum well can easily
be made larger than subband energy splittings, the characteristic energy
for growth direction degrees of freedom.  In very narrow quantum wells,
where higher $T_c$'s may be possible, growth direction degrees of freedom
will be less important.

\subsection{Delta-Doped Layers}

\begin{figure}
\epsfxsize=3.5in
\centerline{\epsffile{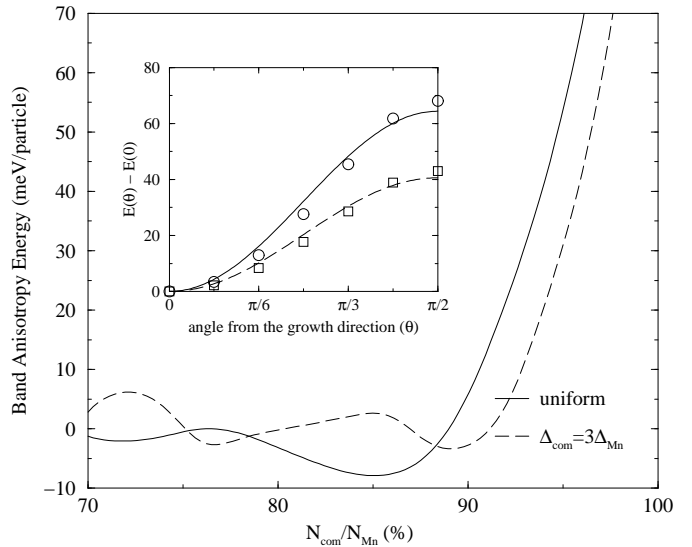}}
\caption{Magnetic anisotropy energy of delta-doped layer. The main plot shows
the dependence of anisotropy energy on carrier density, expressed in
terms of the relative compensation of Mn acceptors. 
The anisotropy energy is due to spin-orbit coupling in the valence band.
The main figure shows the difference between the total energy 
per valence band hole with the magnetization directed in plane and with the 
magnetization in the growth direction.
Assuming nominal compensation, the number of carriers is the difference between 
the number of Mn and twice the number of antisites.
The inset shows the energy as a function of the angle between the 
growth direction and the magnetization direction for 96 \% compensation.
In both cases, the solid line is for uniform As antisite distribution and dashed line for the
Gaussian antisite distribution whose width is $3 \times$ the width of Mn 
distribution.}
\label{aniso} 
\end{figure}

Another type of 
quasi two-dimensional DMS ferromagnet with relatively high ferromagnetic 
transition temperatures was realized by Kawakami {\it et al.} and 
Chen et al.\cite{chen}, by growing monolayers of Mn in bulk systems.
Mn is expected to substitute for all group III elements in a single layer,
with some diffusion of Mn into nearby layers likely.  
The growth of these samples is motivated by the expectation that high local 
Mn densities can be established without forming the decoupled Mn compound
precipitates that occur in the bulk at high Mn concentrations.  
The highest ferromagnetic transition temperatures that have been 
achieved in isolated (Ga,Mn)As digital alloy layers are\cite{kawakami} 
$\sim 50{\rm K}$ 
but transition temperatures larger than $\sim 400 {\rm K}$
have been reported\cite{mccombemaut02} in the corresponding (Ga,Mn)Sb systems.  
(It has not yet been determined whether or not the very high apparent transition 
temperatures in the antimonide case are due to the formation of weakly linked 
MnSb islands.) Some insight into the nature of ferromagnetism 
in uniform digital-doped 
layers can be obtained by applying continuum Mn mean-field
theory\cite{joaquin,byounghak} to this case.  
A digital layer of ${\rm Ga}_{0.5}{\rm Mn}_{0.5}{\rm As}$ provides a 2D density of 
acceptors equal to $N^{2D}_{Mn} \approx 3 \times 10^{14} {\rm cm}^{-2}$. 
Holes are electrostatically attracted to the delta-doped layer and, in the continuum
Mn approximation, form quasi-2D subbands.  
The number of occupied subbands and the degree of localization of the 
subband wavefunctions both depend on the number of layers 
the Mn ions spread over and on the spatial distribution of antisites, the
most common defects in low-temperature MBE grown III-V epilayers. 
In Fig.~\ref{aniso} we plot results obtained from self-consistent 
mean-field calculations when it is assumed that the Mn ions are spread over three 
atomic layers with a Gaussian distribution and that the antisite defects are 
spread uniformly over a width of $12 {\rm nm}$ surrounding the delta-doped layer.
A compensation factor similar to the one  measured for the bulk 
suggests an overall 2D density of $\sim 10^{14} {\rm cm}^{-2}$, about
an order of magnitude larger than a largest 2D density that could be completely 
removed with gates in typical structures.  
We find that in this circumstance four 2D subbands are partially occupied, 
with most of the 2D density in bands that have primarily majority spin 
heavy-hole and majority spin light-hole character.  
The magnetic anisotropy energy
is  sensitive to the degree of compensation 
as illustrated in Fig.~\ref{aniso}.   
We remark that long-range ferromagnetic order
can occur at finite temperatures in 2D only when magnetic anisotropy 
creates a magnetic easy-axis.  The scale of the anisotropy energy
per Mn ion that we find is $\sim 20 {\rm meV}$, much larger
than thermal energies at the mean-field critical temperature $T_c$,
demonstrating that delta-doped layer ferromagnets have strong Ising
character.  As indicated in Fig.~\ref{aniso}, however, the 
magnetic anisotropy can accidentally become small at particular 
densities.  

\begin{figure}
\epsfxsize=3.5in
\centerline{\epsffile{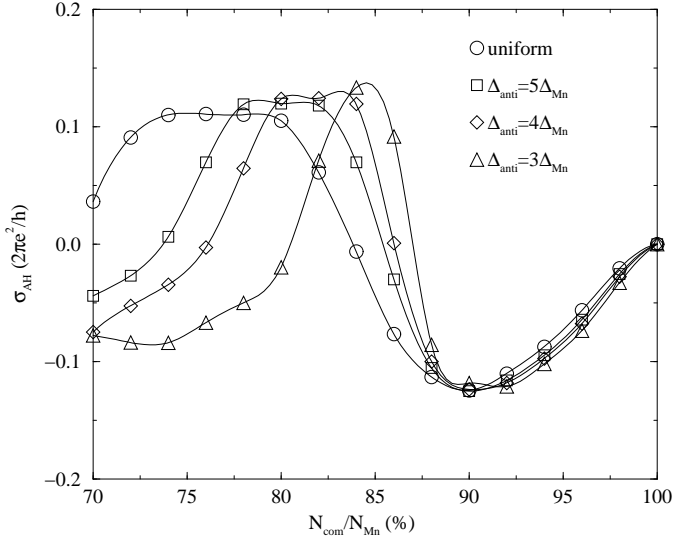}}
\caption{Anomalous Hall conductance in delta-doped layers as a function of
Mn acceptor compensation fraction.  The different symbols correspond to different antisite 
distribution assumptions as in the previous figure.
Gaussian distributions is assumed for both Mn acceptors and antisite defects.}
\label{ahe} 
\end{figure}

The anomalous Hall effect (AHE) in a ferromagnetic metal is commonly used as 
a convenient proxy for the magnetization, especially in quasi two-dimensional
systems for which bulk magnetization measurements may be difficult. 
It has been demonstrated recently\cite{ourahe} that the 
AHE measured\cite{classicrefs,ohnosci98,ohnojmmm99}
in bulk DMS ferromagnets is related to the Berry phase acquired by 
a quasiparticle wavefunction upon traversing
closed paths on the spin-split Fermi surface of a ferromagnetic state.
Theoretical results for the AHE of digital-layer systems are 
summarized in Fig.~\ref{ahe} where we see that the anomalous Hall
conductance depends 
not only on the quasi 2D hole density, but also on the spatial
distribution of compensating defects.  The AHE becomes small in the 
limit of small carrier densities, since all occupied states are 
nearly pure heavy-hole in character and their spinors vary only
weakly with wavevector.  It is also worth noting that the AHE is predicted 
by this theory to have extremely small values in electron modulation doped II-VI
quantum well ferromagnets, because weak spin-orbit scattering in the conduction
band gives quasiparticle spinors that are nearly independent of wavevector and 
have very small Berry phases.  {\em AHE measurements will not be effective in 
screening for ferromagnetism in n-type systems.} 

The mean-field theory of critical temperatures in digital layer DMS ferromagnets
was discussed by Fernandez-Rossier and Sham\cite{joaquin}.  
The authors were able to explain experimental observations 
qualitatively, although they used values of $J_{pd}$ substantially larger than 
those currently thought to be appropriate.  
This partial discrepancy might reflect inadequacies of mean-field 
theory in quasi-2D systems, but could also be due to the substantial 
importance of the antisite defect distribution.  
If these are spread out over a much larger volume than the Mn acceptors, 
holes are attracted more strongly to the Mn layers, producing higher 
mean-field transition temperatures.  The average hole density in 
$\delta$-doped systems is comparable to that of bulk DMS ferromagnets, 
moving the $J_0$ bound beyond the critical temperatures predicted by mean-field theory,
making the relationship between carrier density and critical temperature more
complex.  Still we expect that any strategy that increases the hole density near 
Mn ions will increase $T_c$.  It seems possible that it will ultimately be
possible to increase the $T_c$'s of $\delta$-doped layer systems by tailoring the 
distribution and density of antisite defects, and by adding heterojunction 
barrier layers to force carriers toward the magnetic ions.

\subsection{Superlattices} 

DMS ferromagnet superlattices that mimic metallic giant magnetoresistance
multilayers have been grown by several groups\cite{superlattice}, with the most 
recent results reported by Sadowski {\it et al.}\cite{sadowski}.  The latter authors find 
that ferromagnetism can occur in systems with magnetic layers as thin as $\sim 2.2 {\rm nm}$.
In general these superlattice systems have Mn densities in the magnetic layers 
comparable to those of bulk samples.  We would therefore expect comparable critical 
temperatures, as long as the holes do not leak out significantly beyond the Mn containing layers.
The hole density distribution should fall to zero over a distance comparable to the bulk
Fermi wavelength unless additional acceptor doping has been introduced in the non-magnetic
layers.

\begin{figure}[b]
\epsfxsize=3.5in
\centerline{\epsffile{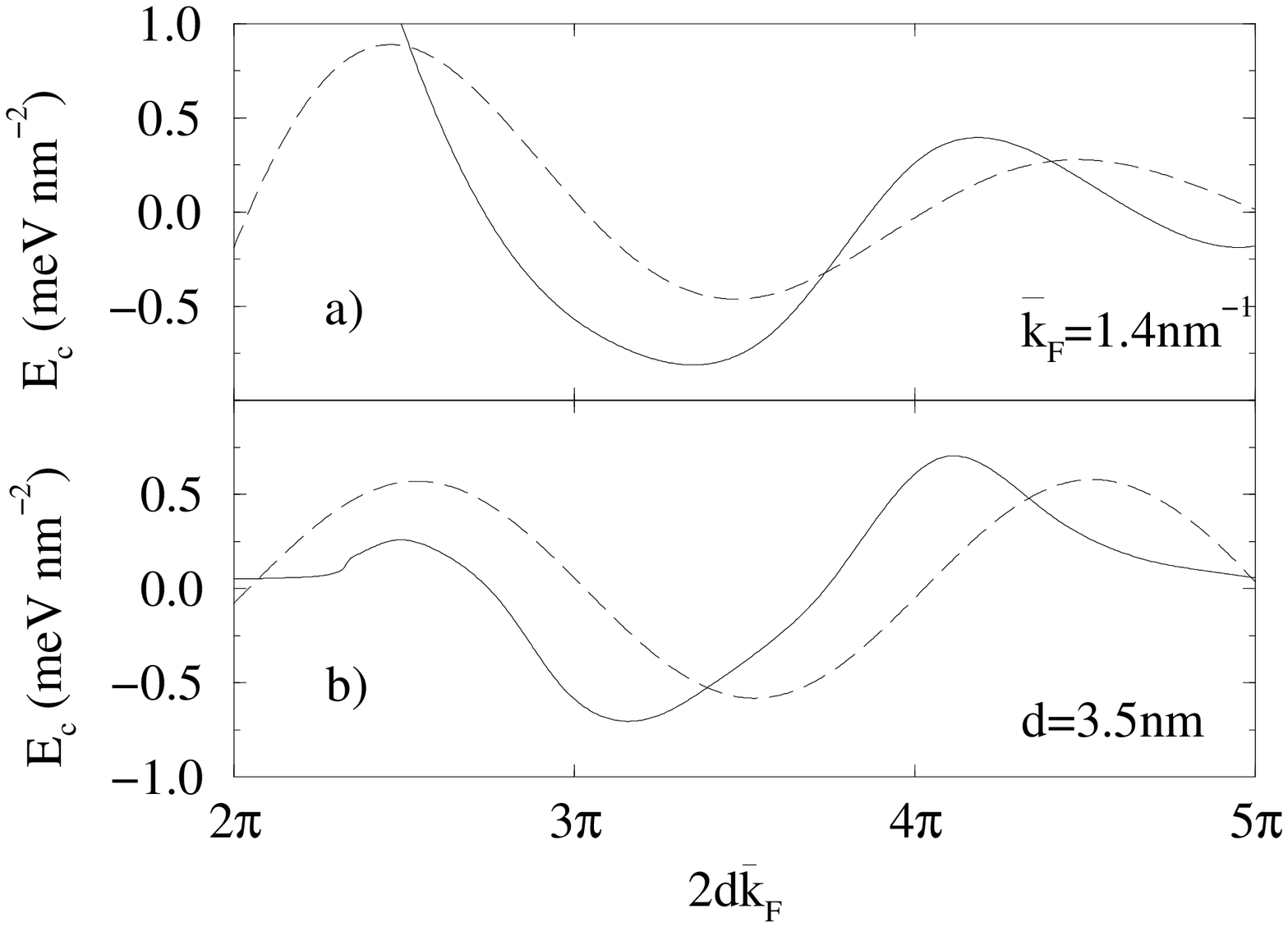}}

\vspace*{.0cm}

\caption{Interlayer exchange coupling (solid lines)
as a function of the GaAs spacer thickness that is free of 
Mn. in (a) and as a function of the average density of free-carriers in (b).
The dashed lines are the RKKY interaction coupling energy
calculated with all magnetic impurities confined to planes
separated by the superlattice period.
Results are plotted as a function of dimensionless parameter
$2d\overline{k}_F$.  These curves were calculated, assuming that 
the Mn acceptor density in the (Ga,Mn)As layers was matched by the 
density of a non-magnetic acceptor in the non-magnetic layers.  
A lower acceptor density, or the absence of acceptors, in the 
non-magnetic layer would produce electrostatic attraction between
holes and magnetic layers and result in weaker magnetic coupling between layers.
}
\label{ec}
\end{figure}

A key property of giant magnetoresitance multilayers is an oscillatory dependence of
the exchange coupling between magnetic layers on the thickness of the non-magnetic spacer.
To illustrate that a similar effect may occur in DMS ferromagnet superlattices, we 
have considered a (Ga.Mn)As/GaAs superlattice with 2~nm thick magnetic layers, $N_{Mn}=1$~nm$^{-3}$, 
and a homogeneous distribution of ionized impurities that neutralize the
free-carrier charge.\cite{ourmftpaper} We look for two different mean-field  solutions, 
a ferromagnetic (F) one with parallel ordered moments
in all Mn-doped regions, and a solution with
an antiferromagnetic (AF) alignment of adjacent magnetic layers. The interlayer
exchange coupling $E_c$, is defined as the difference in energy between
AF and F-states per area
per (Ga,Mn)As layer. 
In Fig.~\ref{ec}  we present numerical results, obtained assuming parabolic bands with effective
mass $m^{\ast}=0.5m_0$, for $E_c$ as a function
of a dimensionless parameter $2d\overline{k}_F$. Here $\overline{k}_F$
is the Fermi wavevector corresponding to the average 3D density of free carriers
in the superlattice with a period $d$. 
Oscillations in the mean-field
(solid lines) $E_c$ are
qualitatively consistent with simple RKKY model estimates.\cite{rkky}
The amplitude of oscillations in $E_c$
is $\sim 10\times$ smaller than in metallic systems\cite{ecmetal}
measured in absolute units and $\sim 10\times$ larger if energy is
measured relative to the Fermi energy of free carriers.
In order to achieve substantial exchange coupling in experimental
systems it will be important to achieve a fairly uniform profile of
ionized acceptors and to limit disorder scattering.  If acceptors are present
only in the magnetic layers, we expect that their exchange coupling 
will be much weaker. 

\begin{figure}[b]
\epsfxsize=3.5in
\centerline{\epsffile{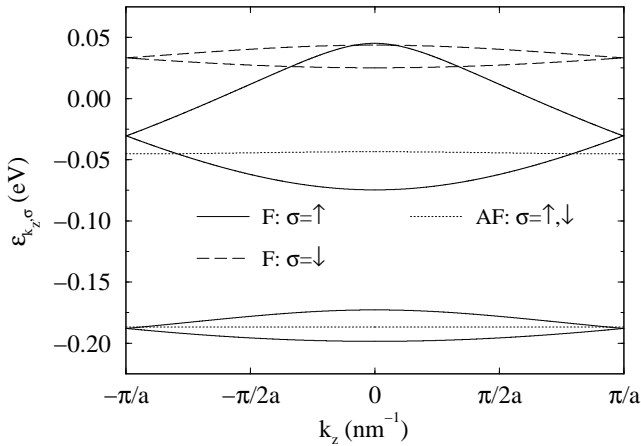}}

\vspace*{.0cm}

\caption{Partially occupied energy superlattice minibands in the 
ferromagnetically (F) aligned superlattice state for spin-up (solid line)
and spin-down (dashed line).  In the antiferromagnetic (AF) superlattice state, 
both spins (dotted line) have the same minibands.  The chemical potential is 0.053~eV and
$a$=7~nm is the unit cell length.  The weak miniband dispersion in the AF case 
would yield the exceptionally strong giant magnetoresistance effects that we 
expect to be typical of layered DMS ferromagnets. 
}
\label{e}
\end{figure}

Fig.~\ref{e} shows occupied minibands
in the superlattice Brillouin zone.
The miniband dispersion is
much weaker in the AF case because the barriers separating
two adjacent minima in the effective potential are twice as thick and
high as in the F case.
Since the conductance is approximately proportional to the square
of the largest miniband width in either coherent or incoherent
transport limits, the minibands can be used to estimate
the size of the current-perpendicular-to-plane (CPP) magnetoresistance
effect.  For the case illustrated, the AF state CPP conductance
will be three orders of magnitude smaller than the F state CPP
conductance. The large difference
is expected since the
bulk (Ga,Mn)As bands are half-metallic
for these parameters.  In general, mean-field theory predicts 
strong CPP magnetoresistance in DMS superlattices if the AF state can be realized.

In closing we remark that envelope-function modelling will fail in very short
period
superlattices.  Vargaftman and Meyer\cite{vurgaftamprb01} 
have recently reported on calculations
using an effective bond
orbital method that is still accurate in this limit.  Their approach will be
useful,
particularly for addressing the possibility of enhancing critical temperatures
by surrounding narrow magnetic layers by barrier material with large band 
offsets.

\section{Concluding Remarks} 

Ferromagnetism in DMS occurs because of exchange interactions between
magnetic ion local moments and itinerant hole or electron spins.
As in other metallic ferromagnets, collective behavior of 
many degrees of freedom results in transport and optical 
properties that can be very sensitive to the ferromagnet's 
magnetic configuration.  This sensitivity is often strongest 
and most easily exploited in trilayer structures in which 
two ferromagnetic layers with weak magnetic coupling are separated 
by an insulating or metallic layer.
DMS ferromagnets differ fundamentally from transition metal ferromagnets because
the density of itinerant carriers is much smaller, 
less than $10^{21} {\rm cm}^{-3}$.  This lower carrier density, combined with  
the essential role that itinerant carriers play in the ferromagnetic 
coupling mechanism, causes magnetic and other properties of 
DMS ferromagnets to be sensitive to defect densities and distributions,
to intentional doping, to strain, and to gates.  Their have recently been a number of 
indications that room temperature ferromagnetism can occur in DMS 
systems, although important properties of these high $T_c$ systems
like homogeneity, carrier-density, and sensitivity to defect 
densities, have not yet been thoroughly explored.  If the same 
degree of malleability already established in lower ($~\sim 100 {\rm K}$) 
$T_c$ DMS ferromagnets can be extended to higher temperatures, 
these ferromagnets are likely to be useful, possibly for spintronics 
applications which currently use transition metals or for  
creative new ideas that integrate information processing
and information storage functions. 

Quantum engineering in semiconductor heterojunction systems 
is more interesting than in metallic systems because of the great flexibility for 
adjusting effective band Hamiltonians on super atomic but still characteristic 
length scales, such as the average distance between carrier.
In this article we have summarized some of the experimental and 
theoretical work, performed mostly over the past five years, on 
semiconductor heterojunction systems containing DMS ferromagnets.  
Although we have made an effort to describe the most important 
experimental and materials developments, 
we have found it convenient to structure our discussion loosely 
around theoretical speculations, predictions, and proposals 
we have made over the past few years, mostly using simple mean-field theory ideas.
It is easier to roam rapidly over the vast phase space of heterojunction
possibilities theoretically than experimentally and, although 
our ability to make confident predictions of magnetic properties is
still limited, we believe that the exercise is useful at this stage of 
the subject.  We have commented here on the limitations of mean-field theory, especially
for predicting the critical temperatures of quantum well DMS ferromagnets,
and emphasized the importance of achieving relatively high carrier densities
in order to get high transition temperatures.  

Research on DMS ferromagnet heterojunction systems is just beginning.
Although our crystal ball is foggy, and likely has a few serious scratches on its
surface, we believe that we can see a few 
enduring truths and principles through the mists of uncertainty. 

\begin{itemize}

\item  It will not be possible to achieve high ferromagnetic transition 
temperatures in modulation doped II-VI quantum well DMS ferromagnets.

\item Magnetic anisotropy is a key property of any magnetic material.
DMS ferromagnet thin films with valence-band hole quasiparticles 
will always be hard magnetic materials because of strong spin-orbit
interactions in the valence band, {\it i.e.}
magnetostatics will play a relatively weak role in determining energetically 
preferred magnetic configurations.  Magnetic anisotropy will
be stronger in quasi two-dimensional p-type DMS ferromagnets and decrease in 
strength with the number of occupied subbands.  n-type DMS ferromagnets,
like those that could occur in modulation doped II-VI quantum well
systems, will be very nearly isotropic purely two-dimensional ferromagnets,
of which their are few examples in nature, and should be of some interest
from a fundamental physics point of view.

\item Itinerant holes or electrons are full participants in the ferromagnetism
of DMS systems.  Band quasiparticle energy splittings at low temperatures 
are comparable to the Fermi energy, lacking complete spin polarization
primarily because of strong spin-orbit interactions in the valence band. 
It will therefore be relatively easy to achieve excellent spintronics functionalities
in spin-valve, magnetic tunnel junction, spin-injector, and other geometries 
using DMS ferromagnets. 

\item The anomalous Hall effect (AHE) will be exceptionally strong in p-type
DMS ferromagnets and exceptionally weak in n-type DMS ferromagnets.
Ferromagnetism in n-type systems is unlikely to be detected by AHE 
measurements.  

\item Quasiparticle state spin-splitting due to magnetic order 
in quasi-two-dimensional DMS ferromagnets
is comparable to the quantum confinement subband splitting in a 10 nm quantum well.   
The interplay between magnetic configurations and growth direction
confinement will be responsible for new physics in quasi-2D 
DMS ferromagnets.

\end{itemize} 

\noindent 
Material science, experimental, and theoretical work is currently in progress
on these and many other related issues.

Our work on DMS ferromagnet heterojunctions has been supported by 
DARPA/ONR Award No. N00014-00-1-095, NSF grant DMR-0115947,
the DOE, the Welch Foundation, the Indiana 
21st Century Fund, the Grant Agency of the Czech Republic, and by the Ministry
of Education of the Czech Republic.  We have benefited from helpful discussions 
with many colleagues especially Ramin Abolfath, Bill Atkinson, David Awschalom, David Baxter, 
Jose Brum, Tomasz Dietl, Dimitri Culcer, Joaquin Fernandez-Rossier, Jacek Furdyna, 
J\" urgen K\" onig, Jan Ku\v{c}era, Hsiu-Hau Lin, Hong Luo, 
Bruce McCombe, Jerry Meyer, Qian Niu, Hideo Ohno, Nitin Samarth, Jairo Sinova, 
Peter Schiffer, John Schliemann, and S.-R. Eric Yang.

%%%%%%%%%%%%%%%%%%%%%%%%%%%%%%%%%%%%%%%%%%%%%%%%%%%%%%%%%%%%%%%%%%%%%%%%%%%%%%
 
\end{document}